\documentstyle [12pt,epsf] {article}
\begin{document}
\bibliographystyle{unsrt}

\pagestyle{empty}               
	
\rightline{\vbox{
	\halign{&#\hfil\cr
	&RHIC Detector Note\cr}}}

\rightline{\vbox{
	\halign{&#\hfil\cr
	&January 1996\cr}}}
\vskip 1in
\begin{center}
{\Large\bf
{Hadronic Spin Dependence and the Use of Coulomb-Nuclear
Interference as a Polarimeter}}
\vskip .5in
\normalsize
T.L.\ Trueman \footnote{This manuscript has been authored
under contract number DE-AC02-76CH00016 with the U.S. Department
of Energy.  Accordingly, the
U.S. Government retains a non-exclusive, royalty-free license to
publish or reproduce the published form of this contribution, or
allow others to do so, for U.S. Government purposes.}\\
{\sl Physics Department, Brookhaven National 
Laboratory, Upton, NY 11973}
\end{center}
\vskip 1.5in
\begin{abstract}
Coulomb-nuclear interference in the single transverse spin
asymmetry $A_N$ is often considered as a possible absolute
polarimeter for proton beams. The main uncertainty in this is the
unknown hadronic spin-flip amplitude. This uncertainty is analyzed
here in the context of the challenge of a $5\%$ polarization
measurement at RHIC. Possible constraints on the spin-flip
amplitude from measurements of the differential cross-section and
the double transverse spin asymmetry $A_{NN}$ are discussed.
\end{abstract}
\vfill \eject \pagestyle{plain}
\setcounter{page}{1}

The addition of polarized proton beams to the RHIC facility
presents the opportunity for an important new and unique physics
program. In order to carry through this program, it is necessary to
have a measurement of the beam polarization. How precise this
measurement must be is not certain, but the challenge of a $5\%$
measurement has been made. A number of possibilities to do this are
under discussion. A very attractive possibility is to make use of
the Coulomb-nuclear interference (CNI) in the single transverse
spin asymmetry $A_N$, which is enhanced in the small $t$ region : $|t|
\sim 10^{-3} GeV^2$ \cite{Lapidus, Buttimore}. This method relies on the
assumption that the hadronic amplitude is spin-independent for small $t$. In
that case, the asymmetry is due solely to the interference between the
electromagnetic spin-flip amplitude and the hadronic non-flip
amplitude, which is determined by the proton-proton total
cross-section. Hence $A_N$ can be calculated and the measurement of
the left-right asymmetry with transversely polarized protons, which is equal to
$P A_N$, would be an absolute measurement of the beam polarization
$P$. The precision of the knowledge of $P$ would be determined by the
precision of the asymmetry measurement and the accuracy of the calculation of
$A_N$.

Unfortunately, there is very little known about the hadronic spin-flip
amplitude in this small $t$ region---indeed, such a measurement is one of
the goals of the RHIC program \cite{Guryn}. Experimentally, there is a fairly
recent measurement at $200 GeV/c$ from Fermilab, but it is not nearly precise
enough to meet the challenge here, and the energy dependence is also
unknown \cite{exp,Akchurin}. Theoretically, so far as I know, there is no
reliable means of calculating to the necessary accuracy the spin dependent
amplitudes for small
$t$. There is an extensive Regge pole fit to low energy data, at $6 GeV/c$
and $12GeV/c$ lab momenta and for $t$ outside the CNI region \cite{Berger}; it
indicates a rather small hadronic spin flip amplitude, but it is large enough
that one would have to correct for it to obtain a 5\% measurement of $P$. In
order to use CNI as a tool for measuring $P$ it is vital that one have a
determination of the spin flip amplitude to the required precision that one
can confidently use in the RHIC energy region. Therefore, in order to reach
some conclusions about the demands of this method, I have adopted a natural and
simple parametrization which has been used before, for example in a proposal
to accelerate polarized protons at Fermilab
\cite{Krisch}. We will see that, in spite of the enhancement of the CNI, one
must know the hadronic spin-flip quite well, roughly to the precision required
for $P$. We will see also that it is difficult to obtain this information
independently from the measurement of the differential cross-section or from
the double transverse spin asymmetry
$A_{NN}$.

For completeness and the convenience of the reader, we will summarize
here the basic formulae required. These are taken from the comprehensive
paper by Buttimore, Gotsman, and Leader \cite{Buttimore}. For proton-proton
scattering, there are five independent amplitudes which will be specified here
in terms of the helicities of the initial and final protons. Conventionally,
these are identified in the following way:
\begin{eqnarray}
\Phi_1(s,t) & = & \langle ++|M|++ \rangle \\
\Phi_2(s,t) & = & \langle ++ |M|-- \rangle \\
\Phi_3(s,t) & = & \langle +- |M|+- \rangle \\ 
\Phi_4(s,t) & = & \langle +- |M|-+ \rangle \\
\Phi_5(s,t) & = & \langle ++ |M|+- \rangle .
\end{eqnarray}
We will decompose each of these amplitudes $\Phi_i$ into a part called $f_i$
due to single photon exchange and a part called $h_i$ due to hadronic
interactions. The $f_i$ are well known and given by
\begin{eqnarray}
f_1(s,t) = f_3(s,t) = \frac{\alpha s}{t} g_D(t)^2 \\
f_2(s,t) = -f_4(s,t) = \frac{\alpha s}{4m^2}  (\mu -1)^2 g_D(t) ^2 \\
f_5(s,t) = - \frac{\alpha s}{2m\sqrt{-t}} (\mu -1) g_D(t)^2, \; \mbox{and} \\
g_D(t) = \frac{1}{(1-t/.71)^2},
\end{eqnarray}
to order $\alpha$. We have systematically neglected the proton mass $m$ with
respect to
$\sqrt{s}$, the center-of-mass energy and we have neglected the momentum
transfer
$\sqrt{-t}$ with respect to $m$. $g_D(t)$ is the dipole form factor of the
proton and $\mu - 1 = 1.79$. The fact that $f_5$ is less singular than
$f_1$ or $f_3$ by a factor $\sqrt{-t}$ is a kinematic effect resulting
from from angular momentum conservation: when $t \rightarrow 0$ the angular
momentum is carried solely by the protons' helicities. It is for the same
reason that $f_4$ has a relative factor of $t$. These are absolute truths
and will be properties of the full amplitudes $\Phi_i$. The remaining
equalities are in a sense accidental and need not be shared with the
amplitudes $h_i$. In practice, lacking any better knowledge, one often
assumes $h_1(s,t) = h_3(s,t)$ and the other $h_i$ to be zero. For example,
in the determination of the total cross-section from the differential
cross-section and the optical theorem this is normally done, although it
may not be correct. 

The issue before us here is not very sensitive to any of these
assumptions except that regarding $h_5$ and to that it is very sensitive.To
see this, let us write down the expressions for the measurables in the same
approximations as above. 
First the total cross-section is given by

\begin{equation}
\sigma_{tot} = \frac{4\pi}{s}Im(\Phi_1(s,t) + \Phi_3(s,t))|_{t=0} ;
\end{equation}
note that this is not effected by the singularity of $f_1$ at $t=0$. Second,
the differential cross-section is given by 
\begin{equation}
\frac{d\sigma}{dt} = \frac{2\pi}{s^2} (|\Phi_1|^2 + |\Phi_2|^2 + |\Phi_3|^2 +
|\Phi_4|^2 + 4|\Phi_5|^2).
\end{equation}
The single and double transverse spin asymmetries are given by
\begin{equation}
A_N \frac{d\sigma}{dt} = -\frac{4\pi}{s^2} Im\{\Phi_5^*(\Phi_1 + \Phi_2 +
\Phi_3 -\Phi_4)\},
\end{equation}
and
\begin{equation}
A_{NN} \frac{d\sigma}{dt} = \frac{4\pi}{s^2} \{2|\Phi_5|^2 + Re(\Phi_1^*
\Phi_2 -
\Phi_3^* \Phi_4) \}.
\end{equation}
Strictly speaking $A_{NN}$ is quite sensitive to all of the $\Phi_i$, but
for the point that we wish to make the assumptions that $h_1 = h_3$ and that
$h_2 = - h_4 = 0$ will not be important and we will adopt these
conventional assumptions in order to focus on $h_5$.

In order to have a reasonable parametrization of the sensitivity, lacking
any better knowledge, we will assume \cite {Krisch}
\begin{equation}
h_5(s,t) = \tau \, \sqrt{-t/m^2} \, h_1(s,t),
\end{equation}
and, for small $t$, we assume the form
\begin{equation}
h_1(s,t) = s \, \frac{(\rho + i)}{8\pi} \, \sigma_{tot} \, e^{bt/2} .
\end{equation}
We use values estimated to be appropriate for RHIC energy \cite{parameters}: $b
= 15 GeV^{-2},\\
\sigma_{tot} = 62 mb \; \mbox{and} \; \rho = 0.13$. This parametrization is
consistent with everything we know and so it represents a possible spin-flip
amplitude. It
does not seem to be a very strong assumption, provided we confine it to a small
range of $t$. Note that $\tau$ is, in general, complex (and a function of
$s$). With these forms $A_N$ has no explicit $s$-dependence, but there is
significant $s$-dependence coming from the $s$-dependence of $\sigma_{tot}$ and
$\rho$. In particular, the height of the peak in $A_N$ decreases 
with energy, roughly as $1/\sqrt{\sigma_{tot}}$ and the $t$-value of the peak
moves toward zero, as
$1/\sigma_{tot}$. 

Because of this form the interference between $h_5$ and $f_1$ has the same
shape in $A_N$ as the CNI interference between $h_1$ and $f_5$ so they just
combine linearly, in the present approximation. If $\tau$ is real, then $h_1$
and $h_5$ are in phase and there is no purely hadronic spin-flip; the shape of
$A_N(t)$ is thus very sensitive to $Im\tau$ as $t$ increases {\em beyond the
CNI region.} Explicitly,
\begin{equation}
A_N \frac{d\sigma}{dt} = \frac{\alpha \sigma_{tot} e^{bt/2}}{2 m \sqrt{-t}}
\{(\mu - 1) - 2 Re(\tau) - 2 \rho Im(\tau) \} + 2 Im(\tau)
\frac{\sqrt{-t}}{m}\left(\frac{d\sigma}{dt}\right)_{hadronic}
\end{equation}
\begin{figure}
\centerline{\epsfbox{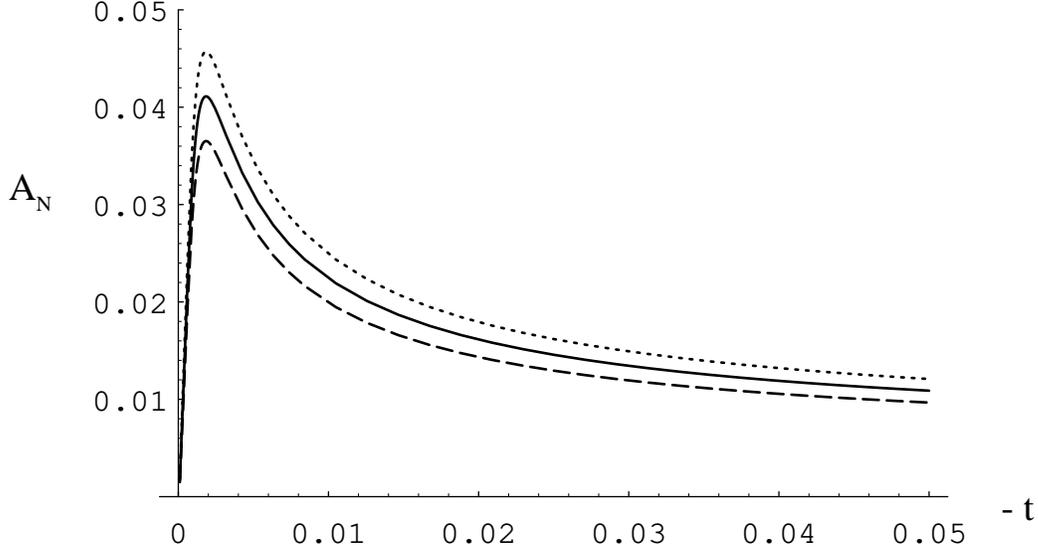}}
\medskip
\caption{$A_N$ for $\tau = 0 \; \mbox{solid curve}, \tau = -0.1 \;
\mbox{dotted}, \tau = +0.1 \; \mbox{dashed}$}
\end{figure}

The main point of this exercise is made by Fig.1. Here we have plotted $A_N$
for three real values of $\tau$: $\tau = 0$, which is the pure CNI case, and
for comparison curves for $\tau = \pm 0.1$ This doesn't seem like an
unreasonably small value of $\tau$ and in assessing the usefulness of this
method, it must be considered in the absence of other constraints. It is
apparent from Fig.1 that all three curves have very similar shape and so, if
for example one assumed that the solid curve were correct when in fact the
dotted curve is correct, one would overestimate $P$ by about $10\%$. Because
the shapes are the same over the entire $t$-range there is no way to separate
$P$ and $A_N$ in the asymmetry measurement. To emphasize this, in Fig.2 we
plot the ratios of the two curves with non-zero $\tau$ to the pure CNI curve
over this $t$-range.
\begin{figure}
\centerline{\epsfbox{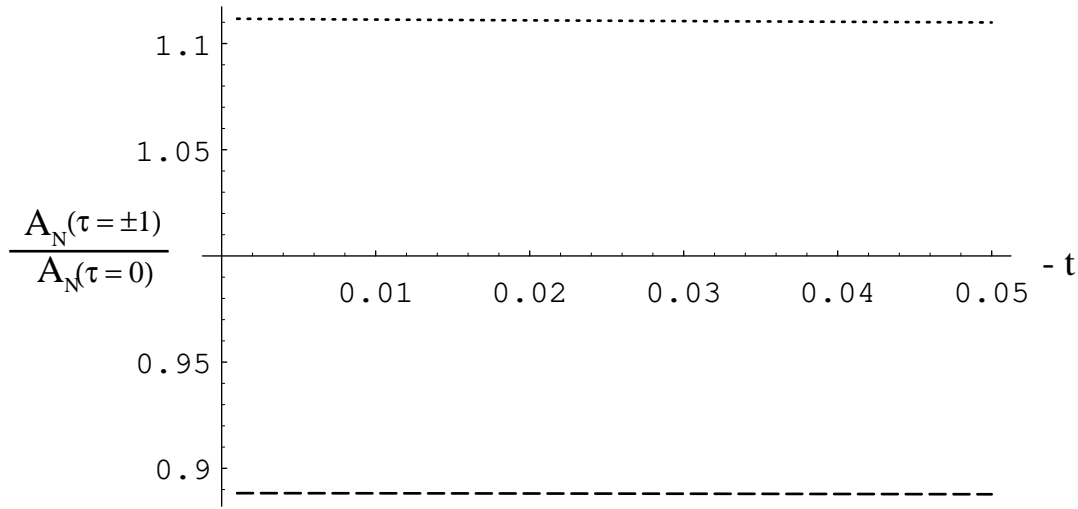}}
\medskip
\caption{The ratio of $A_N$ for $\tau \pm 0.1 $ to the pure CNI case, $\tau =
0$}
\end{figure}
We see that these ratios are constant to a very high degree of accuracy.

If $\tau$ has a non-zero imaginary part, even a very small one can cause a
significant modification of the shape of the $t$-dependence of $A_N$, because
this leads to a purely hadronic contribution which grows, in this
parametrization with $|t|$. One might hope to use this shape to get a handle
on the size of $h_5$. However the real and imaginary parts are independent
and so one cannot really make use of this. This is illustrated in Fig. 3 and
Fig.4, in which $\tau$ is taken to have small positive and negative imaginary
parts, respectively. The three curves in each case correspond to the same
values for $Re \tau$ as in Fig.1; the imaginary parts are nearly the same for
each of the curves in each figure, slightly adjusted to keep the ratios (as
in Fig.2) nearly constant over the whole $t$-range. (This required
fine-tuning for Fig.4 to ensure that the three curves cross 0 at the same
value of $t$, but that is just cosmetic.) The point is that, although one can
readily tell that there is deviation from pure CNI present, because the
curves are in constant ratio one cannot separate $P$ from $A_N$ in these
cases either.
\begin{figure}
\centerline{\epsfbox{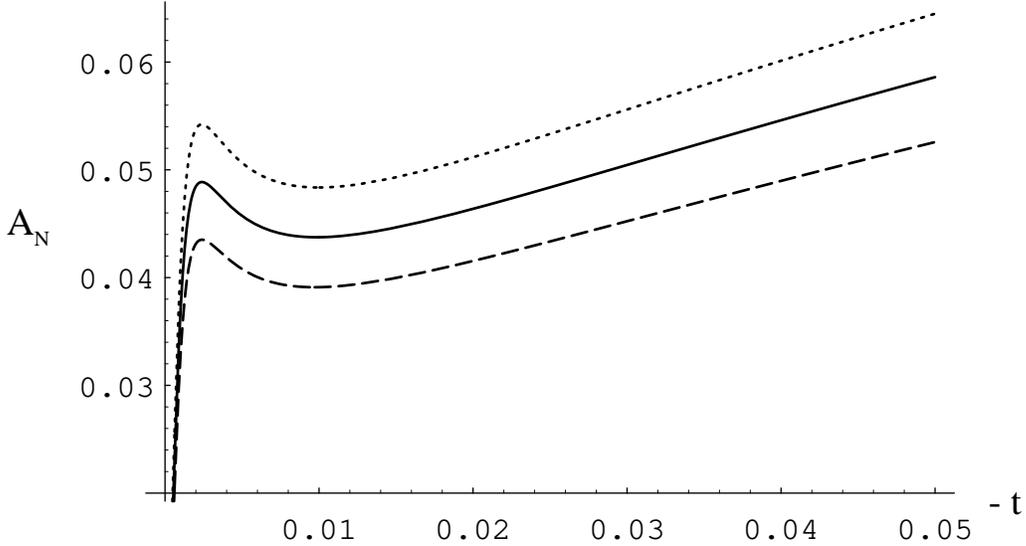}}
\medskip
\caption{$A_N$ for $\tau = 0.1i \; \mbox{solid curve}, \tau = -0.1 + 0.11i \;
\mbox{dotted}, \tau = +0.1 + 0.09i \; \mbox{dashed}$}
\end{figure}
\begin{figure}
\centerline{\epsfbox{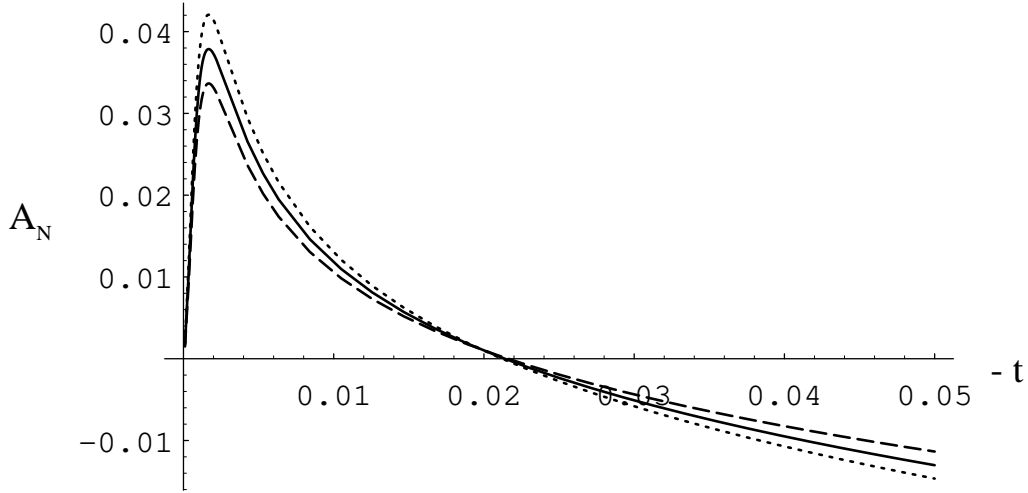}}
\medskip
\caption{$A_N$ for $\tau = -0.05i \; \mbox{solid curve}, \tau = -0.1-0.055i \;
\mbox{dotted}, \tau = +0.1 - 0.045i \; \mbox{dashed}$}
\end{figure}

It is important to see if there is additional information that could be made
available that would constrain $h_5$ to be sufficiently small---or to measure
it sufficiently well---to enable a $5\%$ measurement of $P$ to be made. In
particular, $h_5$ enters into both the differential cross-section and the
double transverse spin asymmetry $A_{NN}$ independently from the way it
enters $A_N$: one enters an unpolarized process and the other
enters a process for which the asymmetry is quadratic in $P$. In principle,
this should be possible. The problem is that in neither one of these is the
contribution of $h_5$ enhanced by interference with the one photon exchange
as it is in $A_N$. 

Consider first the differential cross-section. Buttimore \cite{constraint} has
derived a formula which gives a bound on $h_5$ in terms the differential
cross-section, the total cross-section (presumably measured in an
independent experiment) and the diffraction slope $b$. In the notation of
our {\it ansatz} the bound reads
\begin{equation}
|\tau| < \sqrt{\frac{bm^2}{2} \left(\frac{16 \pi b
\sigma_{el}}{(1+\rho^2)\sigma_{tot}^2}-1\right)}
\end{equation}
The difficulty in using this is that $bm^2 \sim 13$ and so even a bound of
0.1 on $\tau$ would require measuring the combination of quantities in
parenthesis to be equal to 1 to better than one part in $10^3$. This bound
does not seem to provide what is needed.

Finally, let us look at $A_{NN}$. It is useful to look at an approximate
form for it to compare with Eq.(16). Using the same approximations with our 
{\it ansatz} we have
\begin{equation}
A_{NN}\frac{d\sigma}{dt}=
\sigma_{tot} e^{bt/2}
\frac{\alpha}{4m^2} (\mu-1)\{(\mu -1)\rho -4(\rho Re\tau -
Im\tau)\} -2\frac{t}{m^2} |\tau|^2
\left(\frac{d\sigma}{dt}\right)_{hadronic}
\end{equation}
Notice that there is no purely one photon exchange contribution to this
asymmetry either, although at first sight from Eq.13 it might look like there
is. On the other hand the omitted terms containing $\Phi_2$ and $\Phi_4$ could
make significant contributions; their presence would only make using this
asymmetry a more problematic way of constraining $h_5$---we cannot do better
than this. The important point is that the first term in not enhanced by a
$1/\sqrt{-t}$ factor as is the corresponding term in Eq.(16). The second term
is also suppressed by a similar factor. Thus, one is not surprised when one
looks at Fig.5 and sees that $A_{NN}$ corresponding to the parameters in
Fig.1, which translated into a $\pm 11\%$ variation in $P$, is only of order
$0.1\%$. The
corresponding values for the cases of Fig.3 and 4 are only about a factor of 2
larger than this. This, too, seems like a bound that is unlikely to be useful.
\begin{figure}
\centerline{\epsfbox{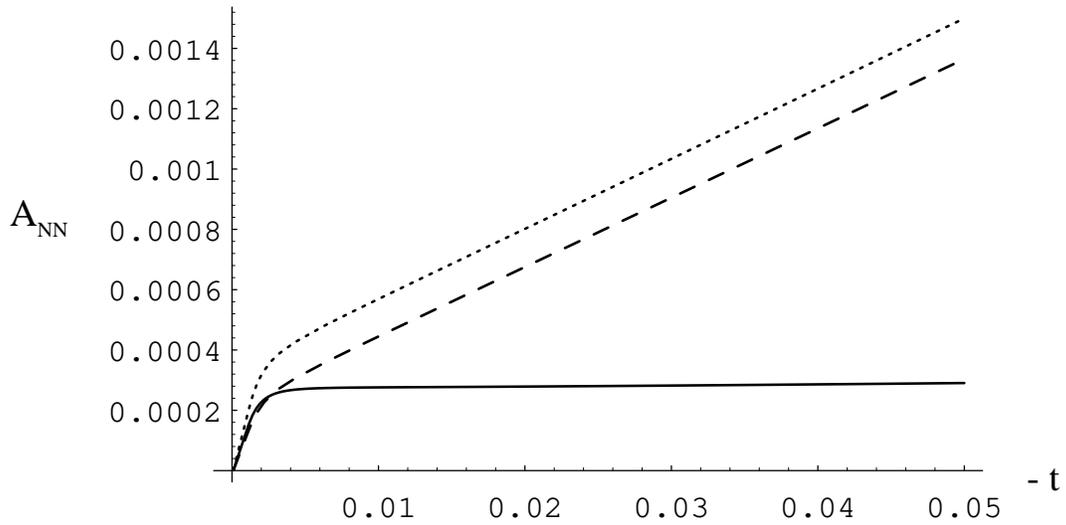}}
\medskip
\caption{$A_{NN}$ for $\tau = 0 \; \mbox{solid curve}, \tau = -0.1 \;
\mbox{dotted}, \tau = +0.1 \; \mbox{dashed}$}
\end{figure}

In conclusion, it is clear from these test cases that the uncertainty in
$h_5$ must be at least as small as the precision required for $P$ in order
for the CNI method to meet the requirement. Neither the $t$-
dependence of $A_N$ nor the measurement of $\frac{d\sigma}{dt}$ or of
$A_{NN}$ are likely to provide the information needed.
 Eventually through a combination
of experiments and theory, we will likely know enough about the spin
dependence of the proton-proton amplitudes to use CNI as an absolute
polarimeter; in the meantime CNI might be used as an indication of the beam
polarization but the measurements must be recognized as provisional and a
source of systematic error. This underlines the importance of measuring the
spin dependence of proton-proton scattering at RHIC.

I would like to thank Y. Makdisi, G. Bunce, J. Soffer, A. Krisch and E. Berger
for valuable discussions.

\end{document}